\documentclass[showpacs,amsmath,prb,preprint]{revtex4-1}

\usepackage[dvips]{color,graphicx}
\DeclareGraphicsExtensions{.eps, .jpg, .png}

\usepackage{graphicx}
\usepackage{dcolumn}
\newcolumntype{d}{D{.}{.}{2}}
\usepackage{bm}
\usepackage{float}
\usepackage{color}
\usepackage{amsmath}
\usepackage{amssymb}

\usepackage{multirow}
\usepackage{url}

\usepackage{pstricks}

\begin{document}

\title{Dimensionalities and multiplicities determination of crystal nets}

\author{Hao Gao$^1$}
\author{Junjie Wang$^1$}
\author{Zhaopeng Guo$^1$}
\author{Jian Sun$^1$}
\email[]{To whom correspondence should be addressed. E-mail: jiansun@nju.edu.cn}

\affiliation{
$^1$National Laboratory of Solid State Microstructures, School of Physics and Collaborative Innovation Center of Advanced Microstructures, Nanjing University, Nanjing 210093, China
}

\begin{abstract}

Low-dimensional materials have attracted significant attentions over the past decade. To discover new low-dimensional materials, high-throughout screening methods have been applied in different materials databases. For this purpose, the reliability of dimensionality identification is therefore highly important. In this work, we find that the existence of self-penetrating nets may lead to incorrect results by previous methods. In stead of this, we use the quotient graph to analysis the topologies of structures and compute their dimensionalities. Based on the quotient graph, we can calculate not only the dimensionality but also the multiplicity of self-penetrating structures. As a demonstration, we screened the Crystallography Open Database using our method and found hundreds of structures with different dimensionalities and high multiplicities up to eleven.

\end{abstract}

\maketitle

\noindent{\bf INTRODUCTION}\\

Low-dimensional materials have been a hot research area in the recent years for their novel properties and potential applications. Some quantum phenomena, such as fractional quantum hall effects\cite{laughlin_anomalous_1983} and Luttinger liquids\cite{luttinger_exactly_1963}, can be realized in the two-dimensional and one-dimensional systems. Due to their novel electronic properties induced by the geometric limit, low-dimensional materials have also been widely applied in various research areas including batteries, catalysis, electronics, and photonics\cite{ferrari_science_2015,bhimanapati_recent_2015}. One way to design new low-dimensional materials is the top-down approach, in which the low-dimensional materials are exfoliated from known bulk phases. Many recent investigations have been focused on searching in large databases for compounds containing low-dimensional components and using high-throughout computational methods to discover new low-dimensional materials with appealing properties\cite{bjorkman_van_2012,lebegue_two-dimensional_2013,mitina_topology_2013,gorai_computational_2016,ashton_topology-scaling_2017,cheon_data_2017,zhang_computational_2018,zhang_effective_2018,mounet_two-dimensional_2018,haastrup_computational_2018,larsen_definition_2019,zhou_2dmatpedia_2019}. For example, previous work have identified thousands of layered structures from more than 100,000 compounds\cite{mounet_two-dimensional_2018}, and about 2,000 of them are exfoliable.

Therefore, a correct and efficient identification of structure dimensionality is highly desired for high-throughout mining of low-dimensional materials. Topology-scaling algorithm (TSA)\cite{ashton_topology-scaling_2017} and rank determination algorithm (RDA) \cite{mounet_two-dimensional_2018} are widely used to determine the dimensionality of a crystal structure. Both methods do not require prior information like the stacking direction for the layers and can dual with complex situations that components with different dimensionalities coexist. However, they require building supercell to connect periodic images. If the supercell is not large enough, TSA and RDA will underestimate the dimensionalities of self-penetrating structures\cite{larsen_definition_2019}. To solve the problem, Larsen \textit{et al.} have proposed a modified RDA method\cite{larsen_definition_2019}. Early works\cite{thimm_graph_2008,blatov_interpenetrating_2004} based on quotient graph\cite{chung_nomenclature_1984} have proposed correct algorithm for dimensionality which has been implemented in ToposPro\cite{blatov_applied_2014} and Systre\cite{noauthor_gavrog_nodate}. Interpenetration analysis is also available through this approach\cite{blatov_interpenetrating_2004,eon_topological_2016}. Quotient graph have also been applied to crystal structure prediction\cite{winkler_systematic_2001,strong_systematic_2004,he_complex_2018,shi_stochastic_2018}, structure decomposition\cite{ahnert_revealing_2017} and machine-learning models for materials property prediction\cite{isayev_universal_2017,xie_crystal_2018,chen_graph_2019,jorgensen_materials_2019}.

Here, we revisited quotient graph (QG) as a powerful tool to discuss dimensionalities and multiplicities of crystals including self-penetrating structures. We demonstrated and discussed a systematical approach based on quotient graph to compute correct dimensionality. The method can deal with the multiplicities of self-penetrating structures correctly as well. Moreover, we mined structures with high multiplicities up to 11 in the Crystallography Open Database (COD)\cite{grazulis_crystallography_2012} to show the reliability of our method.\\

\noindent{\bf RESULTS}\\

\noindent{Definition of quotient graph and component dimensionality}\\

For a crystal structure, the atoms and bonds can be viewed as nodes and edges, and they compose an infinite, undirected graph, called a net. Because of the translation symmetry in the crystal, a net can be described by a finite quotient graph (QG). A QG is a labeled and directed graph containing $N_{at}$ nodes, where $N_{at}$ is the number of atoms in the unit cell. To distinguish the translationally equivalent atoms, we use a notation $n_i(\bm{v})$ ($1 \leq i \leq N_{at}$) to represent the $i$th atom with a Cartesian position $(\bm{x}_i + \bm{v})h$, where $\bm{x}_i$ is the fractional coordination of atom $i$, $h$ is the cell matrix and $\bm{v}$ is an integer vector representing the coordinate of cell. If a bond exists between $n_i(\bm{v}')$ and $n_j(\bm{v}'')$, the corresponding QG has an edge $n_i \xrightarrow{\bm{v} = \bm{v}'' - \bm{v}'} n_j$ labeled by $\bm{v}$. The edge denotes equivalent bonds between $n_i(\bm{v}_0)$ and $n_j(\bm{v}_0 + \bm{v})$ with an arbitrary integer vector $\bm{v}_0$. Obviously, the edge $n_i \xrightarrow{\bm{v}} n_j$ is equivalent to $n_j \xrightarrow{-\bm{v}} n_i$ with the opposite direction.

A connected component $\bm{X}$ in a crystal might contain multiple equivalent atoms \\
$\{n_i(\bm{0}),n_i(\bm{v}_1), n_i(\bm{v}_2), n_i(\bm{v}_3), \dots\}$ and its dimensionality is defined by the rank of the subspace spanned by these connected and translationally equivalent atoms\cite{larsen_definition_2019}:
\begin{equation} \label{eq:dim}
    dim(\bm{X}) = rank(\{\bm{v}_1, \bm{v}_2, \bm{v}_3, \dots\}).
\end{equation}

For extensive (1D, 2D, 3D) components, the set $\{\bm{v}_1, \bm{v}_2, \bm{v}_3, \dots\}$ is infinite, so we should find the basis set before identifying the dimensionality. In this work, a systematical approach based on cycles of QG is used to get the basis set. A cycle is defined as a closed-chain path in a graph. For QG, it represents a path from an atom $n_i(\bm{0})$ to its equivalent atom $n_i(\bm{v})$ in the crystal structure. The relative cell coordinate $\bm{v}$ between the pair of equivalent and connected atoms is equal to cycle sum\cite{thimm_graph_2008} of the cycle. To calculate the cycle sum, for every edge in the cycle, we add the label vector if the direction of the edge is same to that of the cycle, or minus the label vector if the edge direction is opposite. Therefore, the set $\{\bm{v}_1, \bm{v}_2, \bm{v}_3, \dots\}$ in Eq.~\ref{eq:dim} is cycle sums of all the cycles in QG. All cycles in a connected component compose a vector space called cycle space, so we only need to consider cycles in the basis set. Let a matrix $\bm{S}$ consist of cycle sums over the basis set of cycle space, and the dimensionality of the component is
\begin{equation} \label{eq:dim2}
    dim(\bm{X}) = rank(\bm{S}).
\end{equation}\\

\noindent{Previous methods}\\

In TSA, an original cluster in the unit cell contains $N_1$ atoms, then the expanded cluster in an $n \times n \times n$ supercell has $N_2$ atoms. The component dimensionality is determined by the scaling factor $N_2/N_1$. The factor is expected to be $n^d$ ($d=0,1,2,3$), where $d$ is the dimensionality.

RDA is based on the definition of component dimensionality in Eq.~\ref{eq:dim}. However, the original version of RDA\cite{mounet_two-dimensional_2018} used a fixed $3 \times 3 \times 3$ supercell, which is not large enough for complicated structures. In practice, the size of required supercell is unknown in advance. In a modified RDA\cite{larsen_definition_2019}, a breadth-first-search (BFS) is used so that the dimensionality can be determined in a finite number of steps for those components containing infinite atoms.\\

\noindent{Multiplicities of self-penetrating nets}\\

Larsen \textit{et al.}\cite{larsen_definition_2019} have discussed the contrived self-penetrating helical networks and the improper connections between components which lead to incorrect dimensionality by TSA. Here we shall discuss self-penetrating nets following Thimm's approach \cite{thimm_graph_2008} which provides more insights to this problem.

\begin{figure}[htbp]
    \includegraphics[width=0.48\textwidth]{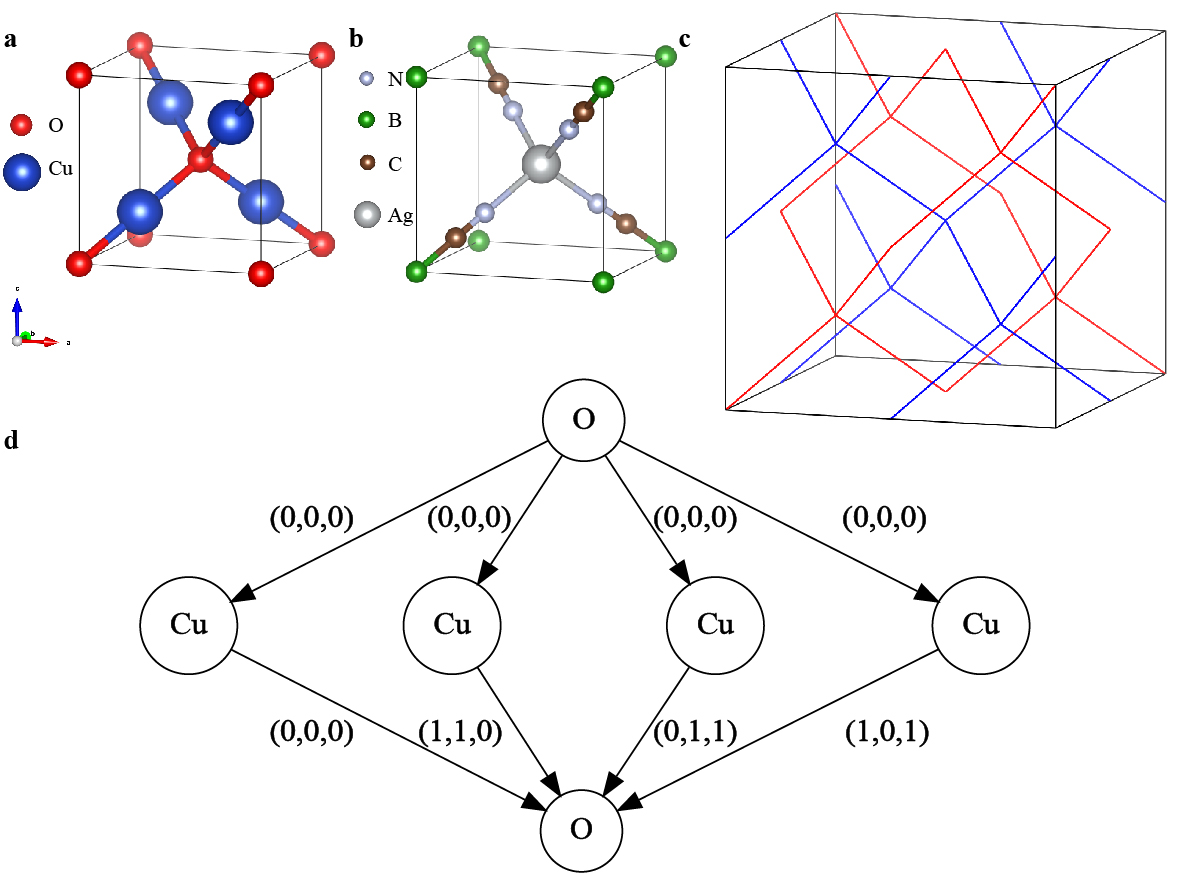}
    \caption{The crystal structures of Cu$_2$O (a) and Ag(B(CN)$_4$) (b). (c) The net with multiplicity of 2 shown in a $2 \times 2 \times 2$ supercell. (d) The QG of Cu$_2$O. The disconnected networks are colored in red and blue.}
    \label{fig:twofold}
\end{figure}

Cuprite with a space group of $Pn\bar{3}m$ is a typical example. As shown in Fig.~\ref{fig:twofold}(a), the O atom with fractional coordinations of $(0,0,0)$ in the original cell $n_{O_1}(0,0,0)$ is connected to the equivalent atoms $n_{O_1}(1,1,0)$, $n_{O_1}(0,1,1)$ and $n_{O_1}(1,0,1)$ through copper and other oxygen atoms. This direct observation can be described by the cycle sum matrix from the QG of cuprite (Fig.~\ref{fig:twofold}(d)):

\begin{equation} \label{eq:cu2o}
  \bm{S} =
  \begin{bmatrix}
   0 & 1 & 1\\
   1 & 0 & 1\\
   1 & 1 & 0
   \end{bmatrix}.
\end{equation}

We found the O atom $n_{O_1}(1,0,0)$ is disconnected from $n_{O_1}(0,0,0)$ because $(1,0,0)$ cannot be represented as a linear combination of three basic vectors in $\bm{S}$ if the coefficients are limited to integers. Actually, cuprite is composed by two disconnected subnets as shown in Fig.~\ref{fig:twofold}(c), and they are equivalent because of the translational symmetry of crystal. Therefore, using TSA, we find one cluster containing 6 atoms ($N_1=6$) in the unit cell and in a $2 \times 2 \times 2$ supercell, the cluster expands to 24 atoms($N_2=24$). The scaling factor $N_2/N_1$ is 4 which leads to an incorrect dimensionality of 2. Another example is Ag(B(CN)$_4$) (Fig.~\ref{fig:twofold}(b)) found by Larsen \textit{et al.}\cite{larsen_definition_2019}, which has the same topology as cuprite.

The net of cuprite contains two translationally equivalent but disconnected subnets, so its multiplicity is 2. For cuprite, the multiplicity $\hat{m}$ equals to the determinant of $\bm{S}$ ($det(\bm{S})=2$)\cite{thimm_graph_2008}. However, in general, the cycle sums matrix is not square. Instead, we should find the basic cycle sums $\tilde{\bm{S}}$. For a 3D net, $\tilde{\bm{S}}$ is a $3 \times 3$ matrix consists of three cycle sums with minimum non-zero absolute determinant among all combinations of cycle sums and the multiplicity equals to the absolute determinant of $\tilde{\bm{S}}$:
\begin{equation} \label{eq:mult}
    \hat{m} = \left | det(\tilde{\bm{S}}) \right |.
\end{equation}
The definition of basic cycle sums for nets with arbitrary dimensionality is proposed by Thimm\cite{thimm_graph_2008}.

\begin{figure}
    \includegraphics[width=0.48\textwidth]{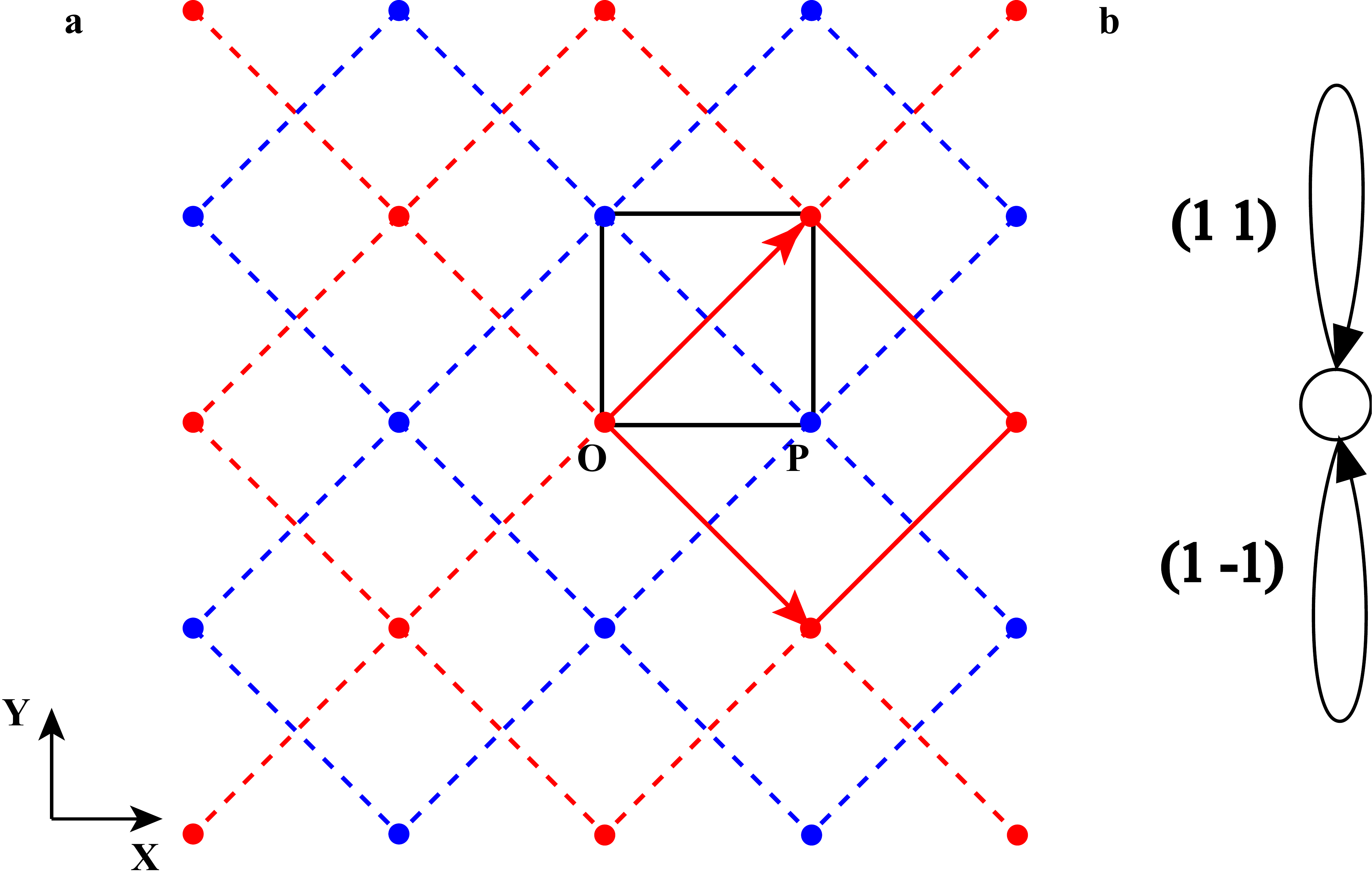}
    \caption{A plane self-penetrating net (a) with multiplicity of 2 and its QG(b). The black box is the unit cell of the square lattice. The red and blue dash lines and points represent the disconnected subnets. The red solid box is the basic building block of the red subnet.}
    \label{fig:plane}
\end{figure}

Thimm has proposed Eq.~\ref{eq:mult} but he has not provided an explanation\cite{thimm_graph_2008}. Here we demonstrate the relation between multiplicity and determinant using a plane self-penetrating net shown in Fig.~\ref{fig:plane}. The basic self-penetrating cell of the red subnet (relative to the unit cell) is defined by the basic cycle sums
$\tilde{\bm{S}} = \begin{bmatrix}
    1 & 1 \\
    1 & -1 \\
\end{bmatrix}$.
The cell contains two points, $O$ and $P$. Since $P$ is in the interior of the cell, its coordinate $(1,0)$ is not an integer linear combination of the basic self-penetrating vectors $(1,1)$ and $(1,-1)$. So it is disconnected from the red subnet. For an arbitrary self-penetrating net, we can always find an $\tilde{\bm{S}}$ to build the basic self-penetrating cell of subnets. Because of the translational symmetry, for each subnet, there is only one point in the basic self-penetrating cell. Therefore, the multiplicity of the net equals to the number of points in the cell, which is the volume/area of the basic self-penetrating cell and the volume/area is $\left | det(\tilde{\bm{S}}) \right |$. We found the basic self-penetrating cell is related to primitive interpenetration cell proposed by Blatov \textit{et al.}\cite{blatov_interpenetrating_2004}

\begin{figure*}
    \includegraphics[width=0.95\textwidth]{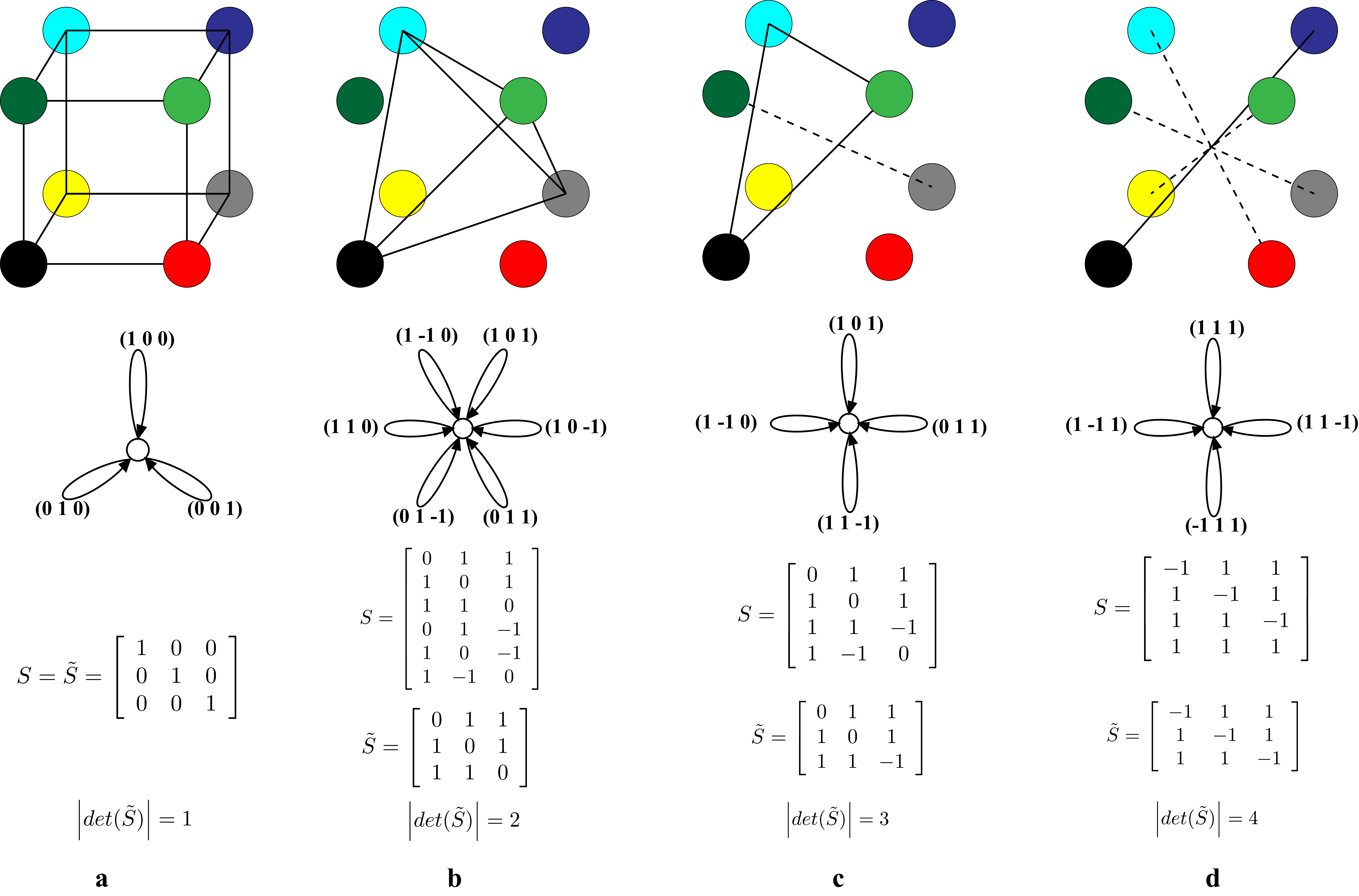}
    \caption{Schematics, quotient graphs, cycle sums and their determinants of 3D nets with multiplicity of 1(a), 2(b), 3(c) and 4(d). The black circle in schematics represents a cluster in the original cell. Other colored circles represent the seven images of the cluster for a $2 \times 2 \times 2$ supercell. The dash lines represent edges disconnected from solid lines.}
    \label{fig:nets}
\end{figure*}

\begin{figure}
    \includegraphics[width=0.48\textwidth]{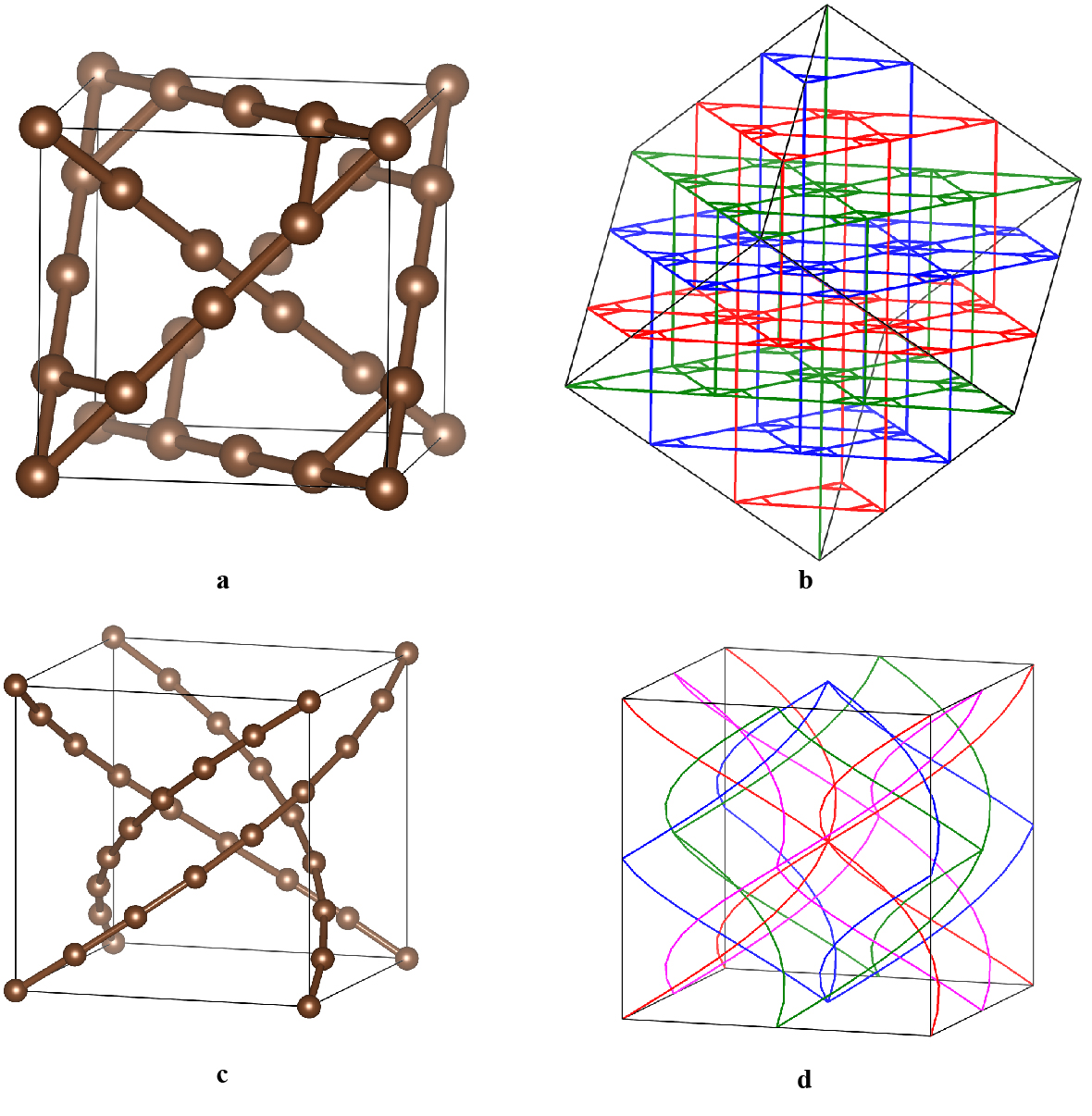}
    \caption{The contrived examples of 3 and 4-fold nets. (a) A 3-fold net in the unit cell. (b) The 3-fold net in a $3 \times 3 \times 3$ supercell. (c) A 4-fold net in the unit cell. (d) The 4-fold net in a $2 \times 2 \times 2$ supercell.}
    \label{fig:34fold}
\end{figure}

In Fig.~\ref{fig:nets}, we have listed examples of 3D net with different multiplicities. The schematics, QGs and cycle sums of a usual 3D net are shown in Fig.~\ref{fig:nets}(a). The original cluster is connected to images in $(1,0,0)$, $(0,1,0)$ and $(0,0,1)$ cell and the multiplicity is 1. In Fig.~\ref{fig:nets}(b), the QG contains six edges which  are along face diagonals in the schematic. It describes nets with multiplicity of 2 like cuprite. Although the QG is different from that of cuprite, their basic cycle sums are similar. Nets with larger multiplicities of 3 and 4 are also possible, as show in Fig.~\ref{fig:nets}(c-d). We can implement edges in the QGs using carbon atomic chains. Multiple disconnected and equivalent components in supercells can be identified, as shown in Fig.~\ref{fig:34fold}. For 4-fold nets, helical atomic chains are used to avoid intersections between edges in the schematic (Fig.~\ref{fig:nets}(d)). TSA also underestimates dimensionalities of these nets. For example, for 4-fold nets, the scaling factor computed by a $2 \times 2 \times 2$ supercell is 2, so the nets are regraded as 1D structure by TSA.

\begin{figure*}
    \includegraphics[width=0.95\textwidth]{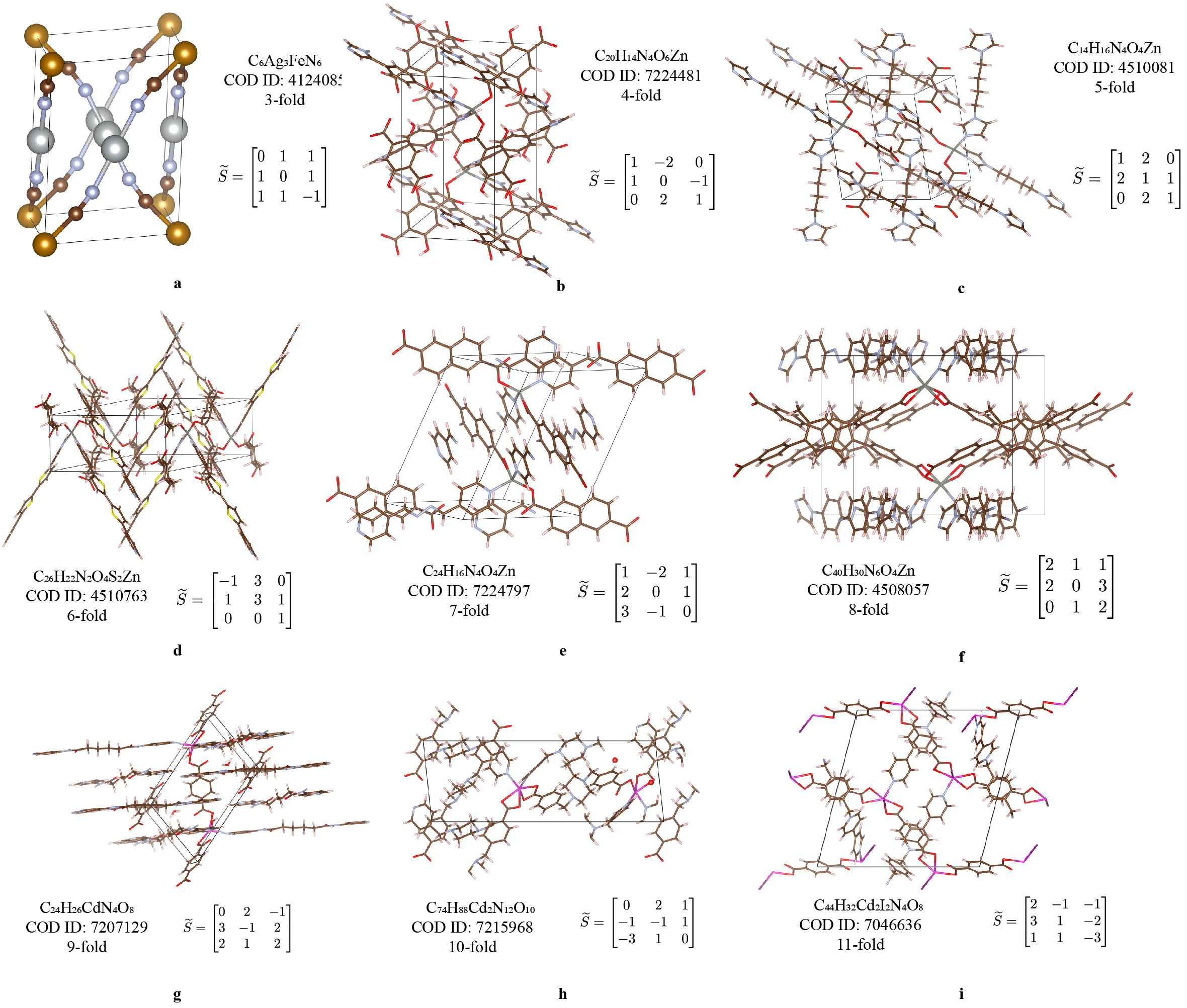}
    \caption{Examples of self-penetrating 3D nets. The crystal structures, chemical formula, multiplicities and basic cycle sums are presented for nets with multiplicities from 3 to 11.}
    \label{fig:3Dnets}
\end{figure*}

Usually, the maximum multiplicity of inorganic 3D nets is four\cite{thimm_graph_2008} and it is related to Hadamard's maximum determinant problem\cite{noauthor_hadamards_nodate}. Hadamard's maximum determinant problem asks for the largest determinant for any $n \times n$ matrix whose elements are taken from a set. For inorganic crystals, the elements of basic cycle sums $\tilde{\bm{S}}$ are usually limited in $\left\{ -1,0,1\right\}$, so $\tilde{\bm{S}}$ is a $(-1,0,1)$-matrix\cite{noauthor_-101-matrix_nodate}. For $n=1,2,3,4,5,\dots$, the largest possible determinant for an $n \times n (-1,0,1)$-matrix \cite{noauthor_-101-matrix_nodate} is $1, 2, 4, 16, 48, \dots$. The sequence is same to maximum multiplicities for $n$-dimensional nets\cite{thimm_graph_2008}. If the elements of $\tilde{\bm{S}}$ are allowed to be larger than 1 or smaller than -1, the maximum multiplicity becomes higher. Such structures are shown in Fig.~\ref{fig:3Dnets} and discussed below.

Both the modified RDA and QG methods can identify correct dimensionalities of self-penetrating nets and QG provides an additional approach to calculating multiplicity. Blatov \textit{et al.} have proposed a general algorithm to compute multiplicities of interpenetrating nets\cite{blatov_interpenetrating_2004,baburin_interpenetrating_2005,carlucci_entangled_2014,v.alexandrov_how_2017}. Actually, the self-penetration discussed here is a special class of interpenetration with only translations.

Based on the database built by Larsen \textit{et al.}\cite{noauthor_definition_nodate}, we have found 3D, 2D and 1D nets with high multiplicities in COD using our method. Here we use interatomic distances to identify the bonds. A bond between atom i and j exists if
\begin{equation} \label{eq:bond}
    d_{ij} < k(r_i^{cov} + r_j^{cov})
\end{equation}
where $d_{ij}$ is the interatomic distance, $r_i^{cov}$ and $r_j^{cov}$ are the atomic covalent radii, and $k$ is the bond-length tolerance parameter. The QGs of crystals highly depend on the value of $k$. For instance, if $k \to \infty$, all structures are identified as 3D and 1-fold nets. Larsen \textit{et al.}\cite{larsen_definition_2019} proposed a scoring parameter to determine the dimensionalities and $k$ intervals and the results are provided in the database\cite{noauthor_definition_nodate}. In this work, for each crystal, we used the low bound for the relative $k$ interval to build the QG and determine the multiplicity. The screening results are shown in Table~\ref{table:netnum}. Note that some nets are self-penetrating only when $k$ is in a narrow interval.

The 3D nets with different multiplicities are shown in Fig.~\ref{fig:3Dnets}. The 3-fold structure, Ag$_3$[Fe(CN)$_6$] (Fig.~\ref{fig:3Dnets}(a)), is similar to the contrived model shown in Fig.~\ref{fig:34fold}(a) since they have same basic cycle sums. Ag$_3$[Fe(CN)$_6$] and the isomorphic compound Ag$_3$[Co(CN)$_6$] have been reported to be colossal thermal expansion materials\cite{goodwin_colossal_2008,goodwin_argentophilicity-dependent_2008}.

For 3D nets in the database, as shown in Table~\ref{table:netnum}, the maximum multiplicity is 11, contrary to the conclusion that the maximum multiplicity of 3D nets is 4. We found that the crystals with multiplicities higher than 3 are all coordination complexes. The long chains allow connections between equivalent atoms in remote cells. Thus, the elements in the basic cycle sums are not limited in $\left\{-1,0,1\right\}$ and the compounds have high multiplicities.

\begin{figure}
    \includegraphics[width=0.48\textwidth]{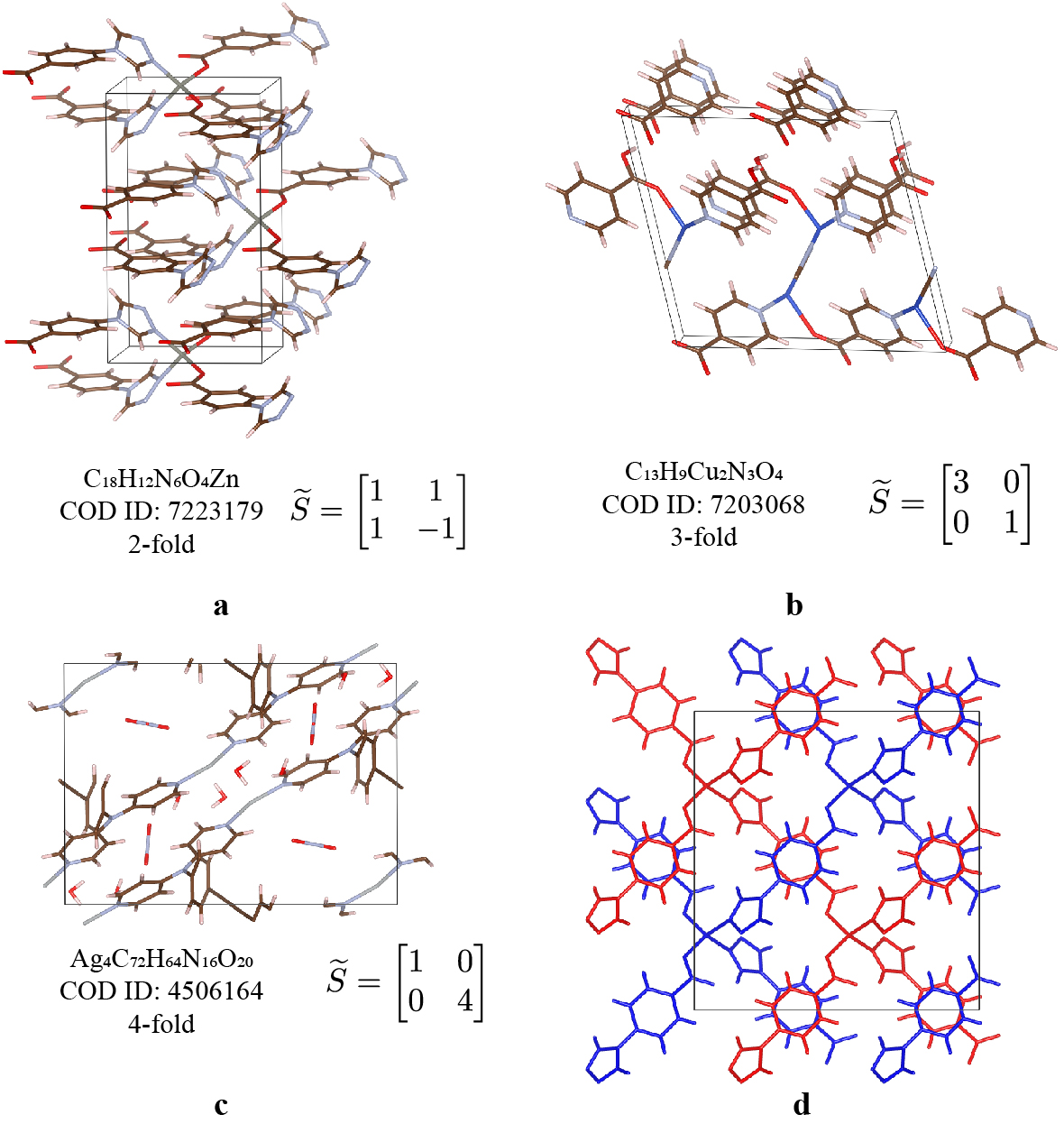}
    \caption{Examples of self-penetrating 2D nets. (a-c) The crystal structures, chemical formula, multiplicities and basic cycle sums of 2D nets with multiplicities from 2 to 4. (d) The 2D 2-fold net in a $1 \times 2 \times 2$ supercell.}
    \label{fig:2Dnets}
  \end{figure}

\begin{figure}
  \includegraphics[width=0.48\textwidth]{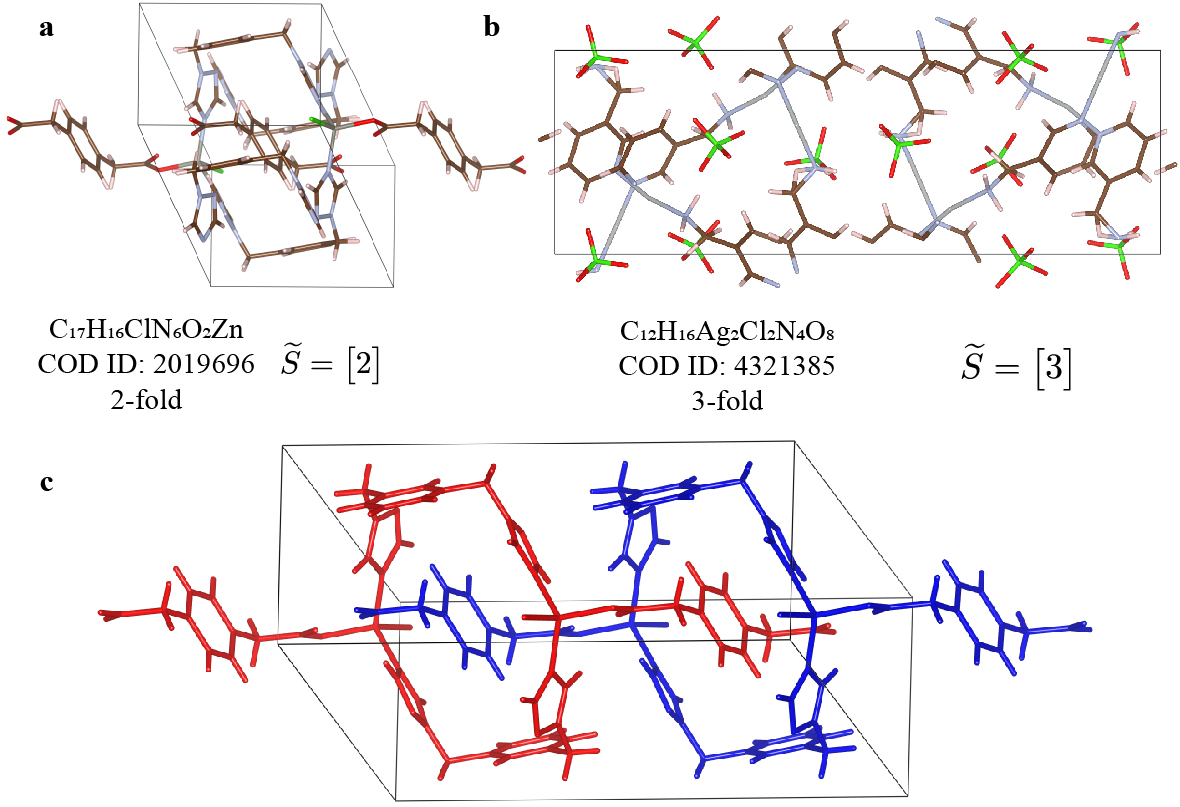}
  \caption{An example of self-penetrating 1D nets. (a-b) The crystal structures, chemical formula, multiplicities and basic cycle sums of 1D nets with multiplicities of 2 and 3. (c) The 1D 2-fold net in a $2 \times 1 \times 1$ supercell.}
  \label{fig:1Dnets}
\end{figure}

We have also found low-dimensional self-penetrating structures shown in Table~\ref{table:netnum}. In a 2D space, we cannot implement a 2-fold net since the edges will always intersect (Fig.~\ref{fig:plane}). But in 3D crystals, atomic chains can curve to form self-penetrating nets. For example, the 2D 2-fold complex (Fig.~\ref{fig:2Dnets}(a)) is similar to the plane self-penetrating net shown in Fig.~\ref{fig:plane}. There are two 2-fold monolayer in the unit cell of the complex and they are stacked along the $a$ axis. In Fig.~\ref{fig:2Dnets}(d), we displayed one of the monolayer in a $1 \times 2 \times 2$ supercell and marked the two components using different colors. Two-dimensional nets with multiplicity of 3 and 4 also exist and the examples are shown in Fig.~\ref{fig:2Dnets}. We have found a 5-fold 2D net but the compound (COD ID: 7216004) is self-penetrating only when $k$ is in a very narrow interval $[1.056, 1.085]$. So it is not regraded as a penetrating polymer in the original reference\cite{ge_assembly_2014}. For 1D structures, self-penetrating nets are very rare and we can only find 2- and 3-fold structures in the database, as shown in Fig.~\ref{fig:1Dnets}. The 1D 2fold complex extends along the $a$ axis and two translationally equivalent components are found to be entangled (Fig.~\ref{fig:1Dnets}(c)).

Mixed-dimensional materials contain multiple components which have different dimensionalities. We have done a statistical analysis on single and mixed dimensionalities in the self-penetrating structure set, as shown in Table~\ref{table:mix}. Mixed-dimensional structures is rare in the whole database\cite{larsen_definition_2019}. However, the proportion of self-penetrating structures in the mix-dimensional set is higher. For instance, 4.9\% of 0D+3D structures are self-penetrating while only 1.4\% of ``pure" 3D structures are self-penetrating.

\begin{figure}
  \includegraphics[width=0.48\textwidth]{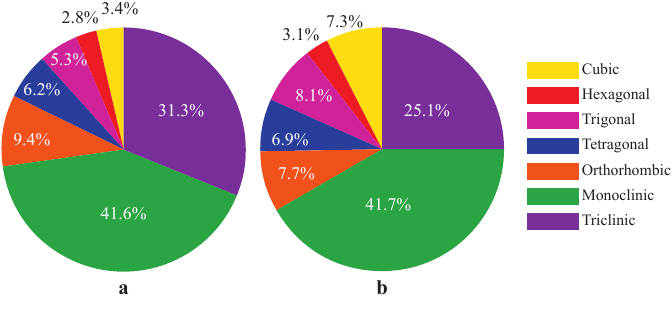}
  \caption{The distributions of crystal systems on (a) all self-penetrating structures and (b) 3D 2-fold structures.}
  \label{fig:stat}
\end{figure}

As shown in Fig.~\ref{fig:stat}, we presented the distribution of crystal systems for the whole self-penetrating set and 3D 2-fold structures, respectively. More than 70\% of self-penetrating structures are triclinic or monoclinic because most of them are organic polymers which have low symmetries. For 3D 2-fold structures, the distribution is roughly similar but the proportion of cubic crystals is much larger. Actually, there are 19 cubic crystals in the self-penetrating set and all of them are 3D and 2-fold. These structures are isomorphic to Cu$_2$O and Ag(B(CN)$_4$) shown in Fig.~\ref{fig:twofold}.\\

\noindent{\bf DISCUSSION}\\

Penetration in materials is usually related to mechanical properties. For example, interpenetrating polymer network (IPN) is a type of elastomer(rubber) which is composed by two or more network polymers\cite{sperling_interpenetrating_1994,roland_interpenetrating_2021}. IPNs based on two polymer materials can improve the mechanical properties like tensile and tear strength\cite{sperling_interpenetrating_1994}. Thus, IPNs have many applications and some commercial materials are IPNs. In recent years, researchers have also proposed new applications of IPNs such as high-performance electroelastomer artificial muscles\cite{ha_interpenetrating_2006}. The self-penetrating polymers screened in this work might have good performances in mechanical properties and wide potential applications. Interpenetration in inorganic atomic networks is highly different from that in polymer materials. Because of the complex interactions in atomic scale, there is no simple relations between interpenetrating and mechanical properties.

In summary, we discussed different dimensionality identification algorithms, such as topology-scaling algorithm and rank determination algorithm. And we found self-penetration in crystal nets will affect the reliability of previous methods. In this work, we introduced a new method to determine the multiplicities of self-penetrating nets by absolute determinant of quotient graph's basic cycle sums. Using this algorithm, we have identified 1D, 2D and 3D self-penetrating crystals in the COD database. Our approach allows for screening structures with target dimensionality and multiplicity in large databases.\\

\noindent{\bf METHODS}\\
We have implemented the algorithms of quotient graph based on Python packages NetworkX\cite{SciPyProceedings_11}, NumPy\cite{walt_numpy_2011} and ASE\cite{larsen_atomic_2017}. We also used ToposPro\cite{blatov_applied_2014} to confirm the results.\\

\noindent{\bf DATA AVAILABILITY}\\
The code and date in this article are available from the corresponding author upon request.\\

\noindent{\bf ACKNOWLEDGMENTS}\\
The authors thank Vladislav A. Blatov for fruitful discussions.
J.S.\ gratefully acknowledges financial support from the MOST of China (Grant No.\ 2016YFA0300404), the National Natural Science Foundation of China (Grant Nos.\ 11974162 and 11834006), the Fundamental Research Funds for the Central Universities.\\

\noindent {\bf AUTHOR CONTRIBUTIONS}\\
J.S. and H.G. designed the study and methodology. H.G. wrote the code and performed the calculations. J.S. and H.G. made the analysis and wrote the manuscript. All authors discussed the results and commented on the manuscript.\\

\noindent {\bf COMPETING INTERESTS}\\
The authors declare no competing interests.


\newpage

\begin{table}[htbp]
    \begin{ruledtabular}
        \caption{Number of nets with different multiplicities and dimensionalities $\hat{m}$ in COD.}
        \begin{tabular}{c c c c c c c c c c c}%
            $\hat{m}$&  2&  3&  4&  5&  6&  7&  8&  9&  10& 11\\
            \hline
            $N_{3D}$&  259&    105& 28& 23& 3&  2&  1& 1& 1& 1\\
            $N_{2D}$&  92&  33& 1& 1\\
            $N_{1D}$&  11& 3\\
        \end{tabular}
        \label{table:netnum}
    \end{ruledtabular}
\end{table}

\begin{table}
    \begin{ruledtabular}
        \caption{Number of entries of each dimensionality type found in the self-penetrating structure set. The percentages are the proportions of self-penetrating structures to all the structures with the dimensionality type. In the diagonal the number of materials with a single dimension are shown while the off-diagonal entries indicate materials with components of two different dimensionalities. In addition to the single- and bi-dimensional materials counted here, we have found a tridimensional 2-fold structures with 0D, 1D, and 3D components(COD ID: 4311765).}
        \begin{tabular}{c c c c c }%
            Dimensions&  0& 1&  2&  3\\
            \hline
            0&  0\\
            1&  8(0.3\%)&  6(0.2\%)&\\
            2&  45(1.8\%)&  0&  82(2.3\%)\\
            3&  112(4.9\%)&  0&  0&  311(1.4\%)\\
        \end{tabular}
        \label{table:mix}
    \end{ruledtabular}
\end{table}

\end{document}